\documentclass[final,3p,onecolumn, 11pt]{elsarticle}

\usepackage{graphicx}
\usepackage{amssymb}
\usepackage{hyperref}

\usepackage{lineno}
\usepackage{amsmath}

\journal{-}

\begin{document}

\begin{frontmatter}


\title{The carbon footprint of distributed cloud storage}



\author[1,2,3]{Lorenzo Posani}
\author[2]{Alessio Paccoia}
\author[2]{Marco Moschettini}

\address[1]{Laboratoire de Physique Statistique, \'Ecole Normale Sup\'erieure, Paris (FR)}
\address[2]{Research and tech development, Cubbit srl, Bologna (IT)}
\address[3]{\normalfont{to whom correspondence should be addressed, email: \href{mailto:me@example.com}{lorenzo.posani@cubbit.io}}}

\begin{abstract}
The ICT (Information Communication Technologies) ecosystem is estimated to be responsible, as of today, for 10\% of the total worldwide energy demand - equivalent to the combined energy production of Germany and Japan. Cloud storage, mainly operated through large and densely-packed data centers, constitutes a non-negligible part of it. However, since the cloud is a fast-inflating market and the energy-efficiency of data centers is mostly an insensitive issue for the collectivity, its carbon footprint shows no signs of slowing down. In this paper, we analyze a novel paradigm for cloud storage (implemented by \href{http://cubbit.io}{\underline{cubbit.io}}) \cite{cubbit, cubbit2}, in which data are stored and distributed over a network of p2p-interacting ARM-based single-board devices. We compare Cubbit's distributed cloud to the traditional centralized solution in terms of environmental footprint and energy efficiency. We demonstrate that, compared to the centralized cloud, the distributed architecture of Cubbit has a carbon footprint reduced of a $\sim 77$\% factor for data storage and of a $\sim 50$\% factor for data transfers. These results provide an example of how a radical paradigm shift in a large-reach technology can benefit both the final consumer as well as our society as a whole.
\end{abstract}


\end{frontmatter}


\section{Introduction}
\label{S:1}

Over the last decades, the general acknowledgment of the climate crisis has driven most western countries towards an increasing awareness of consumptions and efficiency. As a result, increasingly-tight policies led to a consistent decrease of the average per-device consumption of household appliances (fridge, cooling, etc.) \cite{efficiency}. However, there is a largely-underestimated factor contributing to the environmental impact of our daily lives: our use of the Information Communication Technology (ICT) ecosystem or, in other words, our digital life. An estimation of the total impact of the ICT ecosystem approaches 1500 TWh of annual consumption \cite{greenpeace2012cloud,mills2013cloud}, which roughly amounts for 10\% of the world energy consumption, more than the energy production of Germany and Japan combined.

A computation of the per-capita consumption shows that the average use of a personal smartphone is equivalent, without considering the charging costs, to the energy consumption of an additional household fridge \cite{mills2013cloud}. In contrast with the trend of the electronics market, however, the environmental impact of our online life is much less tangible and, as a consequence, much less controversial. The under-estimation the internet's footprint, combined with the fast-increasing trend of online presence and online devices per capita, results in a growing and mostly unopposed environmental impact that shows no signs of slowing down \cite{forecast}. 

Every time a video is streamed from Youtube servers to an iPad, or a photo is accessed on Google Photos or Dropbox, the whole infrastructure that separates the final user from the corporate data center, as well as the data center itself, has to be powered to reliably transmit information in both directions. Depending on the relative location of the exchanging nodes, the process of transmitting information can be orders of magnitude more consuming than storage itself. 

In this document we analyze, using an adaptation of the model of Baliga et al. \cite{baliga2011green}, the energy consumption of \textit{cloud storage} servicess and compare it to an alternative setup where data is stored on peer-to-peer low-consumption devices located in users' houses and implemented by \href{www.cubbit.io}{Cubbit}.

\section{Analysis of centralized cloud consumptions}

\begin{table*}[t]
\begin{center}
    \begin{tabular}{ | l | l | l | l | l |}
    \hline
      Name & Capacity & Wattage (peak) & PUE & Redundancy \\ \hline
      Cubbit Cell & 1.5 TB & 2.55 W & 1.0 & 1.5 \\ \hline
      HP SO 3620 & 96 TB & 607 W & 1.6 & 2.0\\ \hline
      HP SO 5650 & 2240 TB & 6603 W & 1.6 & 2.0\\ \hline
      ECS-D5600 & 2240 TB & 9500 W & 1.9 & 2.0\\ \hline
      ECS-EX300 & 192 TB & 275 W & 1.9 & 2.0\\ \hline
      Storage Pod & 480 TB & 1500 W & 1.6 & 1.1 \\ \hline
    \end{tabular}
\end{center}
\caption{\textbf{Equipments} - power and capacity of storage equipments}
\label{table-specs-storage}
\end{table*}

The energy consumption of a cloud storage service can be divided into two main factors:
\begin{enumerate}
\item the cost of storing the data, i.e. powering and cooling the data center (Storage consumption)
\item the cost of sending the data from the user to the server and back (Transfer consumption)
\end{enumerate}

While the first can be estimated from technical specifications of storage equipment, the second needs a more detailed analysis that takes into account the public internet infrastructure and the geographical distance between the user and the server. For both these estimations we refer to the model of Baliga et al. \cite{baliga2011green}, where energy consumption is computed accounting for several factors, including the multiplicity of involved devices, redundancy, cooling, overbooking (see below).

To delineate the calculation, we start from the storage consumption, i.e. the average power, expressed in W/TB, necessary to store the payload in \textit{hot storage}. We updated the technical specifications with respect to \cite{baliga2011green}, as hard-disk storage capacity has dramatically improved in the last years. 
As a model for data center racks, we considered five of the most recent products from three leading companies in the sector of enterprise storage hardware, see Tab.\ref{table-specs-storage}. For each storage appliance we take the peak consumption, as it is the information made available in the manufacturer specs sheet \cite{hpspecs, ecsspecs, bbspecs}. To estimate the capacity, we consider every rack to be filled with 8TB disks (estimated from the mean of HDD dimensions in a recent BackBlaze report \cite{blackblaze}, $\simeq 7.2 \ \text{TB}/\text{Disk}$) that are fully employed to store customers' cloud files (with no empty space overhead). When the heat produced is not specified in the manufacturer specs sheet (as in the case of HP racks) we consider a 1.6$\times$ cooling factor, estimated by the most recent reports on average Power Usage Effectiveness (PUE) in data centers \cite{}. Similarly, when the redundancy strategy is not explicitly stated, we use a 2$\times$ factor \cite{baliga2011green} as if every files was mirrored two times in the same or in a different data center. Both BackBlaze and Cubbit employ a redundancy protocol based on erasure coding with replication factors $1.1 \times$ and $1.5 \times$, respectively. These values are simply combined in the following formula to obtain the storage consumption per Terabyte:

\begin{equation}
\text{P}^{\text{storage}}_{\text{dcenter}} = \text{PUE} \times \text{redundancy} \times \frac{\text{peak Wattage}}{\text{n disks} \times 8 ~\text{TB}}
\end{equation}

Likewise, we compute the transfer energy, expressed in J/GB, following the public internet model of \cite{baliga2011green}. The analysis relies on the definition of the consumption per bit, which is computed by dividing the operating power (W) by the total transfer capacity (Gb/s), resulting in a Joule/bit measure, then converted in J/GB. These units are taken from the manufacturer specs sheet, shown in table 1. These quantities are then combined with a set of coefficients reflecting the redundancy of the packet transmission, the under-operating regime of the infrastructure and the cooling energy, and the multiplicity of some devices in a single transmission (e.g. two ethernet switches at entry points plus another one inside the data center). The average distance between core routers on the network is estimated to be c.a. 800Km. For a full description of coefficients and estimations we refer the reader to \cite{baliga2011green}.

\begin{table*}[t]
\begin{center}
    \begin{tabular}{ | l | l | l | l |}
    \hline
           & Equipment & Capacity & Consumption \\ \hline
      Data Center gateway router & Juniper MX-960 &660 Gb/s & 5.1 kW\\ \hline
      Ethernet Switch & Cisco 6509 & 160 Gb/s & 3.8 kW\\ \hline
      BNG & Juniper E320 & 60 Gb/s & 3.3 kW\\ \hline
      Provider Edge & Cisco 12816 & 160 Gb/s & 4.21 kW\\ \hline
      Core router & Cisco CRS-1 & 640 Gb/s & 10.9 kW\\ \hline
      WDM (800 km) & Fujitsu 7700 & 40 Gb/s & 136 W/channel\\
    \hline
    \end{tabular}
\end{center}
\caption{\textbf{Equipments} - power and capacity of routing equipments. Data from \cite{baliga2011green}}
\label{table-specs}
\end{table*}

\begin{align}
&\text{E}^{\text{transfer}}_{\text{dcenter}} = \\ \notag = 6 \times &\bigg( 3 \frac{P_{es}}{C_{es}} + \frac{P_{bg}}{C_{bg}} + \frac{P_{g}}{C_{g}} + 2\frac{P_{pe}}{C_{pe}} + 18 \frac{P_{c}}{C_{c}} + 4\frac{P_{w}}{C_{w}} \bigg) \\ \notag &\simeq 23.9~\frac{\text{kJ}}{\text{GB}} \ \ ,
\label{server-transfer}
\end{align}

where the prefactor of $6$ accounts for redundancy ($\times~2$), cooling and other overheads ($\times~1.5$), and the fact that today’s network typically operate at under 50\% utilization ($\times~2$); the addends represent, in order, the ethernet switch, the broadband gateway, the data center gateway, the provider edge router, the core network, and the relay optical fiber transmission. The detailed analysis of pre-factors can be found in \cite{baliga2011green}. Briefly, the factor $3$ in the ethernet switch accounts for the two routers involved in the access to the public internet plus the router located inside the data center; the factor $18$ in the core network accounts for an average of $9$ hops ($2$ baseline + $7$ for the 800km distance between core nodes) of internet packets from source to destination, times $2$ for the redundancy.

\section{Analysis of Distributed Cloud consumptions}
The distributed architecture of the Cubbit network relies on the same public internet infrastructure delineated in the previous chapter. However, the distributed paradigm has three key differences from server-based cloud storage:
\begin{enumerate}
\item The low energy consumption of storage devices (based on the Marvell ESPRESSObin \cite{marvell})
\item The absence of cooling overhead
\item The geographical proximity between the users and their stored data
\end{enumerate}
We consider a network of Cubbit Cells \cite{cubbitio}, each composed by an ARM-based SBC and a HDD (Western Digital Blue) of 1 TB or 2 TB (estimated average 1.5 TB disk). Each Cell is located in a user's house and connected to internet by an internet service provider. Files on Cubbit's cloud are stored with a redundancy factor of 1.5 (Reed Solomon erasure coding with $24+12$ redundancy shards \cite{cubbit, monti2017alternative}). As done for the centralized cloud, we analyze the consumption of both storage (W/GB) and transfer (J/GB).

 The Marvell ESPRESSObin has a single-core peak consumption of $\sim 1$W \cite{marvell}, while the embedded WD Blue HDD has a peak consumption of $1.4$ W (1 TB) and $1.7$ W (2 TB). We here assume that half of the network is composed by 1 TB devices and half by 2 TB devices, giving an average storage of 1.5 TB, corresponding to an average peak consumption of $1.55$W. The storage energy consumption of the Cubbit network is therefore computed as

\begin{equation}
\text{P}^{\text{storage}}_{\text{cubbit}} = 1.5 \times \frac{1 + 1.55~\text{W}}{1.5~\text{TB}} \simeq 2.55~\frac{\text{W}}{\text{TB}} \ \ .
\end{equation}

In Cubbit, shards of the distributed payloads are preferably distributed in Cubbit Cells that are located in geographical proximity of the user, since the distribution of the shards is controlled by the AI optimization routines of a coordinator server\cite{cubbit}. We therefore consider the scenario where data is stored in nodes at an average distance of 80 km from the user's access point. In this scenario, we can assume an average number of 2 packet hops in core network routers. This lowers the corresponding factor $18$ in Eq.~\ref{server-transfer} to a factor $4$, accounting for two core hops and the redundancy of the packets on the network (factor 2). For the same reason, the 800km-relay consumption $P_{w}$ is not taken into account. With respect to Eq.~\ref{server-transfer} we also ignore all data-center specific terms: one ethernet switch and the data center gateway. However, we need to consider an additional BNG, since transfers are performed through p2p connections between endpoints located within an ISP network. The transfer energy per GB is therefore computed as

\begin{align}
\label{server-transfer}
\text{E}^{\text{transfer}}_{\text{cubbit}} &= 6 \times \left( 2 \frac{P_{es}}{C_{es}} + 2 \frac{P_{bg}}{C_{bg}} + 2\frac{P_{pe}}{C_{pe}} + 4 \frac{P_{c}}{C_{c}} \right)\\ \notag &\simeq 11.9~\frac{\text{kJ}}{\text{GB}} \ \ .
\end{align}

\section{Comparison between centralized cloud and Cubbit distributed cloud}
\begin{figure*}[ht]
\centering
	\includegraphics[width=\linewidth]{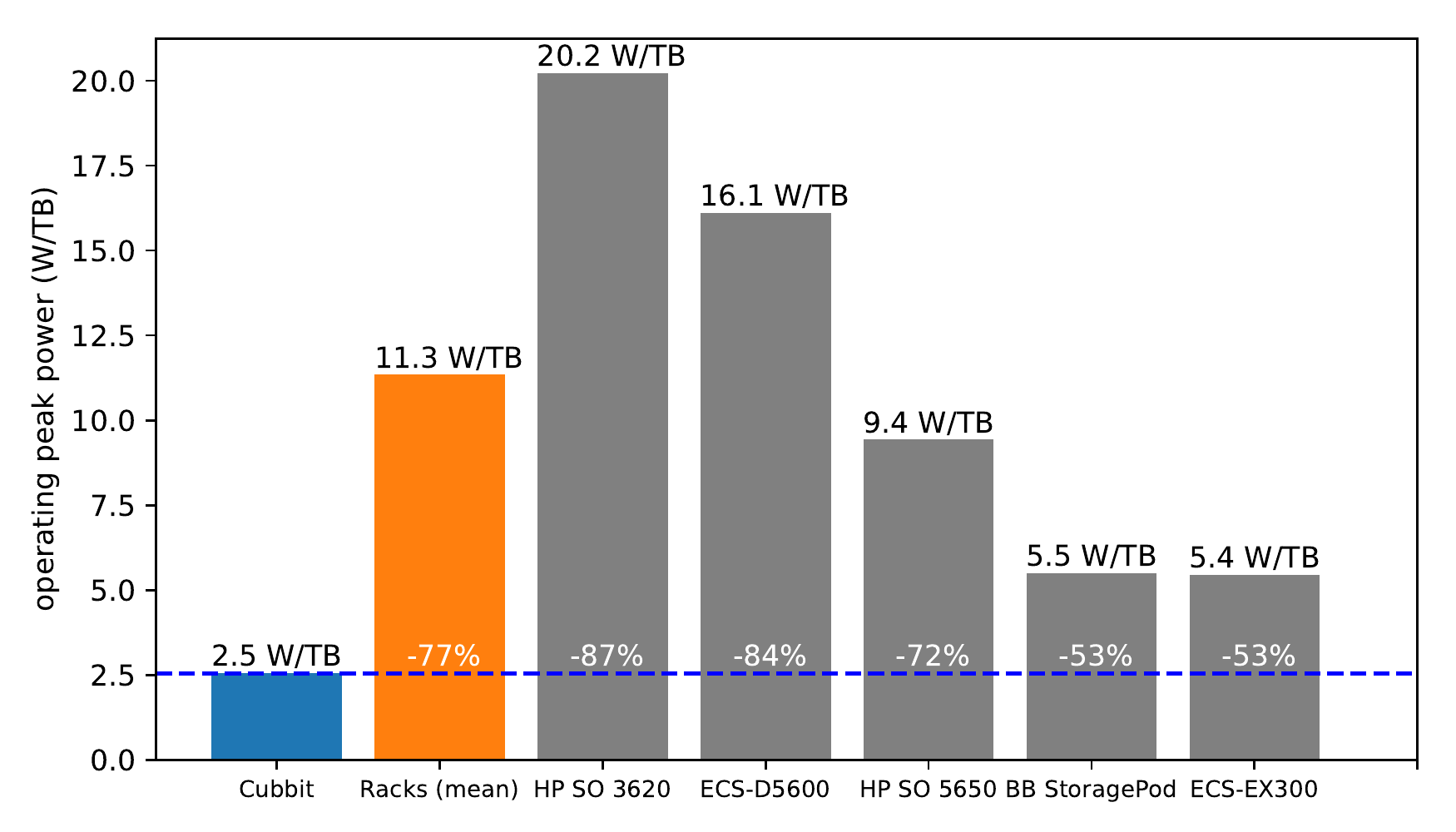}
	\caption{Comparison between centralized and distributed clouds in terms of peak consumption per TB of stored data.}
	\label{compared-storage}
\end{figure*}

As a first analysis, we show in Fig.~\ref{compared-storage} the comparison between Cubbit and the centralized cloud in terms of pure storage consumption. Cubbit achieves a reduction ranging between $\sim 50\%$ and $\sim 95 \%$ with respect to data centers' racks. By taking the average over racks as a mid-range estimation (although we have no information on the relative distribution of these racks in the data center market), we obtain

\begin{equation}
	\Delta P^{\text{storage}} = \left< P^{\text{storage}}_{\text{dcenter}} \right> - P^{\text{storage}}_{\text{cubbit}} \simeq 9 ~\frac{\text{W}}{\text{TB}} \ \ ,
\end{equation}

corresponding to an overall 77\% reduction:
\begin{equation}
	\frac{\Delta P^{\text{storage}}}{\left< P^{\text{storage}}_{\text{dcenter}}\right>} = \simeq 0.77 \ \ .
\end{equation}

Similarly, the difference in terms of transfer energy per GB is

\begin{align}
		\Delta E^{\text{transfer}} &= E^{\text{transfer}}_{\text{dcenter}} - E^{\text{transfer}}_{\text{cubbit}} \\ \notag &\simeq 12.0~\frac{\text{kJ}}{\text{GB}} = 3.33~\frac{\text{kWh}}{\text{TB}}\ \ ,
\end{align}

which corresponds to a 50\% reduction of the energy needed to transfer data from the cloud to the user, and back. \\

The reduction of carbon footprint of Cubbit compared to centralized solutions can be computed by comparing the storage power and the transfer energy for typical use case, such as backup plans and frequent access of, for instance, a web-hosted video.
\subsection{Backup}
A backup service hosted on the cloud is characterized by large volumes that are not frequently accessed. In the context of the carbon footprint, the consumption of a backup plan will, therefore, be dominated by the storage term. If we consider a storage plan for a professional backup of 25 TB with very small daily access, we find that the total energy saved in a year is

\begin{align}
	\Delta &E (\text{25 TB backup}) = \\ \notag &25 \ \text{TB} \times \Delta P^{\text{storage}} \times 365 \times 24 h \simeq 1971 ~\text{kWh} \ .
\end{align}
By considering a rough factor of 0.5 KgCO2 for each kWh of consumed energy\cite{co2}, the use of a distributed cloud over a centralized one would correspond, for such a backup plan, to a reduced carbon emission of c.a. -1000.0 kgCO2/year. If we consider that data centers usually operate on the Petabyte scale, we easily see that the yearly reduction in emissions scales up to hundreds of tons of CO2
\begin{align}
	\Delta &\text{Footprint} (\text{backup}) \simeq  40\ 000 \  \text{kgCO2/year/PB}
\end{align}

\begin{figure}
	\includegraphics[width=\linewidth]{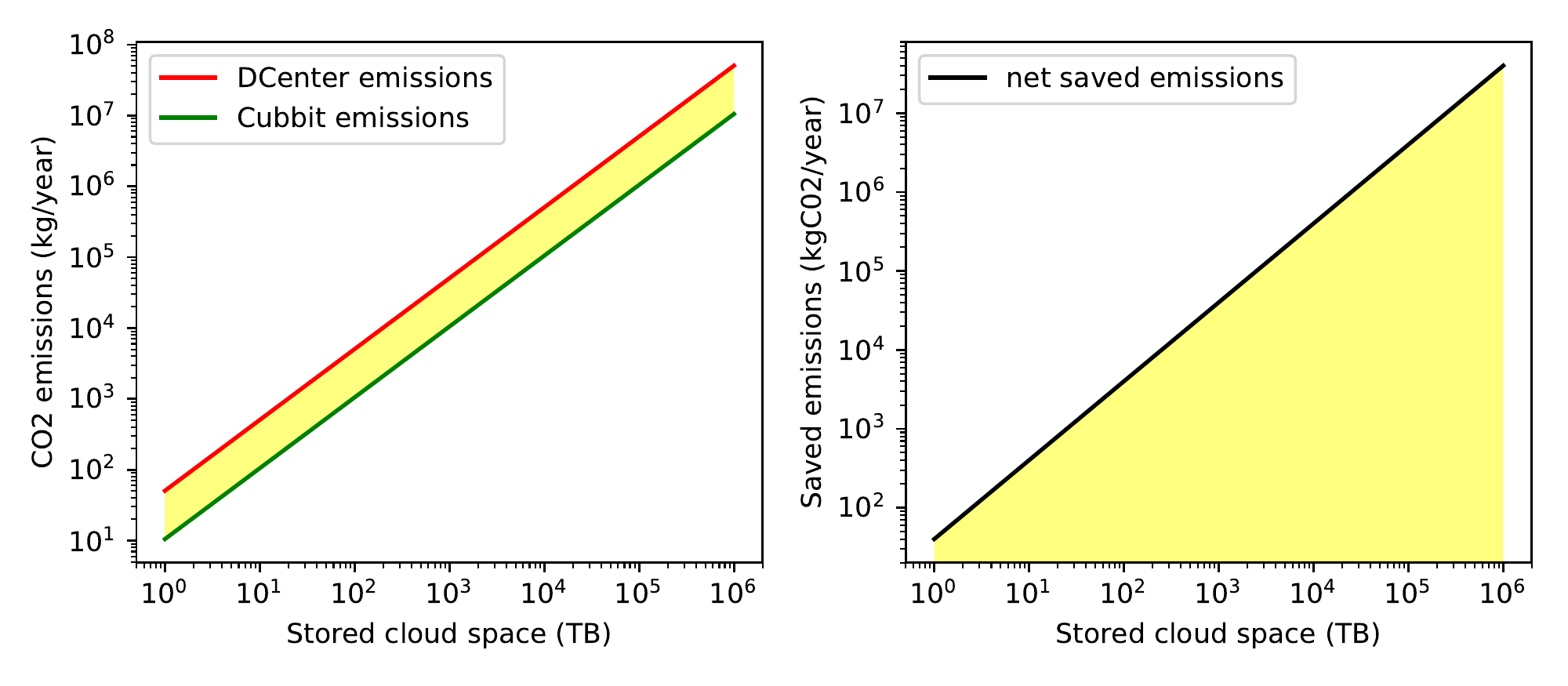}
	\caption{Comparison between centralized (DCenter) and distributed (Cubbit) clouds in terms of annual carbon emissions per TB of stored data.}
\end{figure}

\subsection{Streaming}
The reduction in consumed energy and, consequently, in carbon emission is significantly larger when considering large volumes of data transfers. For example, if we consider a medium news website hosting 25 TB of data and streaming, on average, 10 TB of data per day (e.g. 10,000 visualizations of 100 MB each) the saved energy per year would be

\begin{align}\notag
	\Delta E (&\text{10 TB streaming}) = \\ \notag &25~\text{TB} \times \Delta P^{\text{storage}} \times 365 \times 24 h \\ \notag &+ 365 \times \Delta E^{\text{transfer}} \times 10~\text{TB} \\ &= 14 136 ~\text{kWh} \ ,
\end{align}

which roughly corresponds to -7,000 kgCO2 emitted per year. Note that these computations assume that data are broadcasted to a local audience. While this might be the case for university data, local news, or targeted marketing, it has a limited range of applicability that has to be taken into account.

\begin{figure}
	\includegraphics[width=\linewidth]{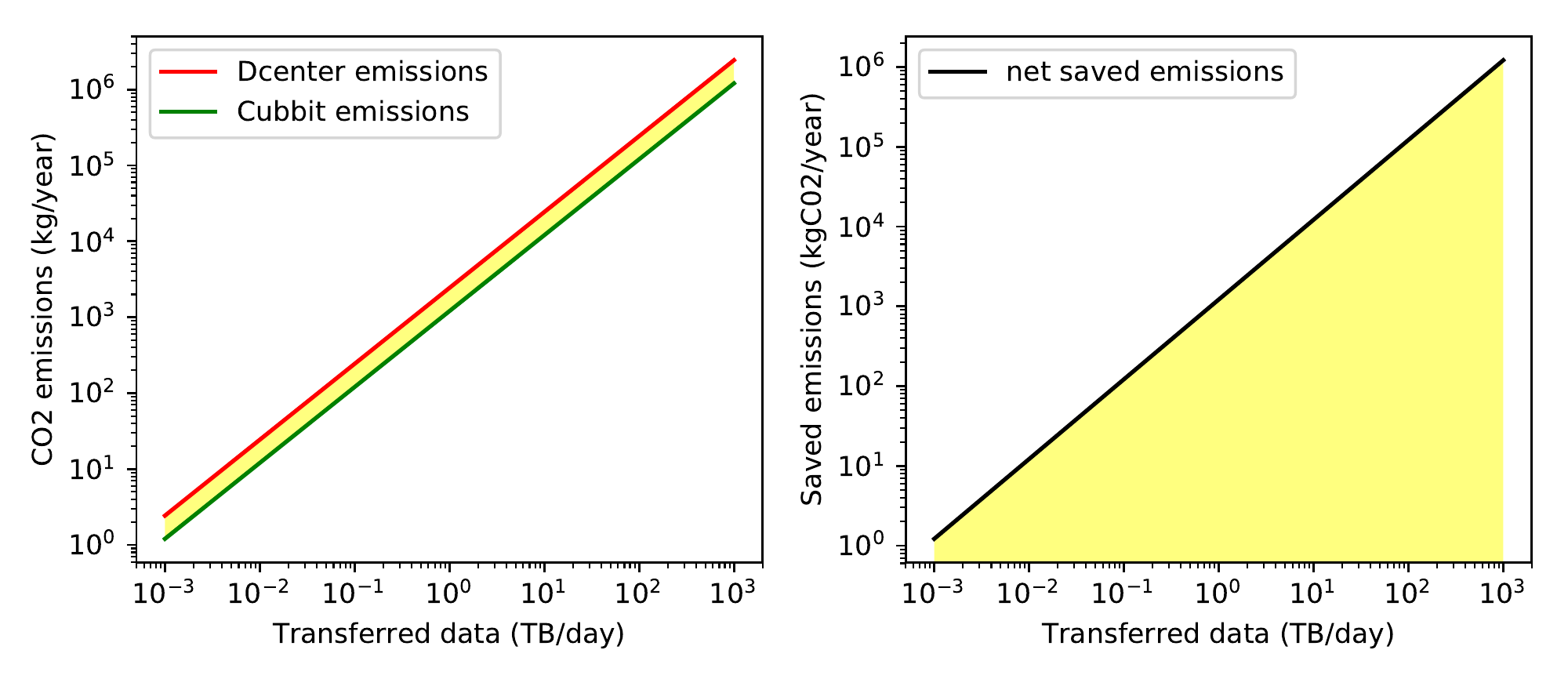}
	\caption{Comparison between centralized (DCenter) and distributed (Cubbit) clouds in terms of annual carbon emissions per TB of daily streamed data.}
\end{figure}
\subsection{Large scale: global cloud industry}

Finally, if we speculate about the overall data volume of a global consumer cloud storage service like, for example, Dropbox or Google, values rise dramatically. Such interpolations have to be taken with due caution, since estimations are based on undisclosed values. For the sake of speculation, we consider a use base of c.a. 600 millions users. The latest disclosed conversion rate from free (2 GB) to premium (2 TB) is around 3\%. This results in a theoretical data volume of ca. $37.2 \cdot 10^6$ TB of storage. Considering a factor 5 due to overbooking, it gives an estimation of $7.4 \ 10^6$ TB of effective cloud storage. We can make a conservative estimation that each user transfers, on average, 50 MB of files from/to the cloud, which implies a daily transfer volume of c.a. 190 TB. If we plug these estimations in our model, we obtain a total saved annual energy, using a distributed architecture rather than a centralized one, of $\sim 6.7 \cdot 10^8$ kWh, equivalent to saving carbon emissions in the order of $300$ million kgCO2 per year.

\section*{Acknowledgements}
The authors are grateful to S. Baldi, A. Albani, and A. Rovai for valuable comments on the manuscript and fruitful discussion.
\clearpage



\bibliographystyle{model1-num-names}
\bibliography{sample.bib}







%
%

\end{document}